\documentclass[%
 aip,
 amsmath,amssymb,
 reprint,%
]{revtex4-1}

\usepackage{graphicx}
\usepackage{dcolumn}
\usepackage{bm}

\usepackage[utf8]{inputenc}
\usepackage[T1]{fontenc}
\usepackage{mathptmx}
\usepackage{etoolbox}

\usepackage{xcolor}

\makeatletter
\def\@email#1#2{%
 \endgroup
 \patchcmd{\titleblock@produce}
  {\frontmatter@RRAPformat}
  {\frontmatter@RRAPformat{\produce@RRAP{*#1\href{mailto:#2}{#2}}}\frontmatter@RRAPformat}
  {}{}
}%
\makeatother
\begin{document}

\preprint{AIP/123-QED}

\title[]{Design and Implementation of Optics for the EXperiment for Cryogenic Large-Aperture Intensity Mapping (EXCLAIM)}
\author{Thomas Essinger-Hileman}%
 \email{thomas.m.essinger-hileman@nasa.gov}
 \affiliation{NASA Goddard Space Flight Center, Greenbelt, MD 20771, USA}



\author{Danny Chmaytelli}
\affiliation{University of California, Los Angeles, Los Angeles, CA 90095, USA}

\author{Trevor Oxholm}
\affiliation{University of Wisconsin-Madison, Madison, WI 53706, USA}
\affiliation{Massachusetts Institute of Technology Lincoln Laboratory, Lexington, MA 02421, USA}

\author{Tatsat Parekh}
\affiliation{NASA Goddard Space Flight Center, Greenbelt, MD 20771, USA}

\author{Gage Siebert}
\affiliation{University of Wisconsin-Madison, Madison, WI 53706, USA}
\affiliation{Arizona State University, Tempe, AZ 85281, USA}

\author{Eric R. Switzer}
\affiliation{NASA Goddard Space Flight Center, Greenbelt, MD 20771, USA}

\author{Joseph Watson}
\affiliation{NASA Goddard Space Flight Center, Greenbelt, MD 20771, USA}

\author{Alyssa Barlis}
\affiliation{NASA Goddard Space Flight Center, Greenbelt, MD 20771, USA}

\author{Emily M. Barrentine}
\affiliation{NASA Goddard Space Flight Center, Greenbelt, MD 20771, USA}

\author{Jeffrey Beeman}
\affiliation{Lawrence Berkeley National Lab, Berkeley, CA 94720, USA}


\author{Christine Chung}
\affiliation{Applied Physics Laboratory, The Johns Hopkins University, Laurel, MD 20723, USA}

\author{Paul Cursey}
\affiliation{NASA Goddard Space Flight Center, Greenbelt, MD 20771, USA}

\author{Sumit Dahal}
\affiliation{NASA Goddard Space Flight Center, Greenbelt, MD 20771, USA}

\author{Rahul Datta}
\affiliation{University of Chicago, Chicago, IL 60637, USA}

\author{Negar Ehsan}
\affiliation{NASA Goddard Space Flight Center, Greenbelt, MD 20771, USA}

\author{Jason Glenn}
\affiliation{NASA Goddard Space Flight Center, Greenbelt, MD 20771, USA}

\author{Joseph Golec}
\affiliation{University of Massachusetts Amherst, Amherst, MA 01003, USA}
\affiliation{University of Chicago, Chicago, IL 60637, USA}

\author{Andrew Lennon}
\affiliation{Applied Physics Laboratory, The Johns Hopkins University, Laurel, MD 20723, USA}

\author{Luke N. Lowe}
\affiliation{NASA Goddard Space Flight Center, Greenbelt, MD 20771, USA}

\author{Jeffrey McMahon}
\affiliation{University of Chicago, Chicago, IL 60637, USA}

\author{Maryam Rahmani}
\affiliation{NASA Goddard Space Flight Center, Greenbelt, MD 20771, USA}

\author{Peter Timbie}
\affiliation{University of Wisconsin-Madison, Madison, WI 53706, USA}

\author{Bruce Tretheway}
\affiliation{Applied Physics Laboratory, The Johns Hopkins University, Laurel, MD 20723, USA}

\author{Carole Tucker}
\affiliation{Cardiff University, Cardiff CF10 3AT, UK}

\author{Carolyn Volpert}
\affiliation{University of Maryland, College Park, MD 20742, USA}
\affiliation{NASA Goddard Space Flight Center, Greenbelt, MD 20771, USA}

\author{Edward J. Wollack}
\affiliation{NASA Goddard Space Flight Center, Greenbelt, MD 20771, USA}


\date{\today}

\begin{abstract}
This work describes the design and implementation of optics for EXCLAIM, the EXperiment for Cryogenic Large-Aperture Intensity Mapping. EXCLAIM is a balloon-borne telescope that will measure integrated line emission from carbon monoxide (CO) at redshifts $z<1$ and ionized carbon ([CII]) at redshifts $z = 2.5-3.5$ to probe star formation over cosmic time in cross-correlation with galaxy redshift surveys. The EXCLAIM instrument is designed to observe at frequencies of $420$--$540$~GHz using six microfabricated silicon integrated spectrometers with spectral resolving power $R = 512$ coupled to kinetic inductance detectors (KIDs). A completely cryogenic telescope cooled to a temperature below 5~K provides low-background observations between narrow atmospheric lines in the stratosphere. Off-axis reflective optics use a $90$-cm primary mirror to provide $4.2^\prime$ full-width at half-maximum (FWHM) resolution at the center of the EXCLAIM band over a field of view of $22.5^\prime$. Illumination of the $1.7$~K cold stop combined with blackened baffling at multiple places in the optical system ensure low ($< -40$~dB) edge illumination of the primary to minimize spill onto warmer elements at the top of the dewar. 
\end{abstract}


\maketitle

\section{Introduction}
\label{sec:intro}
Star formation reached a peak at a redshift $z{\sim}2$ and has subsequently declined despite the continued growth of dark matter haloes.\cite{2020ApJ...902..111W, 2014ARA&A..52..415M} Galactic feedback mechanisms are commonly thought to be the cause of this decline, but further observations are required to disentangle the effects. Line intensity mapping (LIM), in which integrated line emission is mapped at low angular resolution, offers a complementary approach to traditional galaxy surveys with several advantages. LIM creates a map of all emission in a given line free from selection bias. Furthermore, the low angular resolution required enables large survey areas to be efficiently mapped with modest requirements on the telescope aperture size. Simulations show tensions between models and measurements of CO in individual galaxies, motivating a blind, complete survey over a large area.\cite{2019ApJ...882..137P}

The EXperiment for Cryogenic Large-Aperture Intensity Mapping (EXCLAIM) is a balloon-borne cryogenic telescope that will measure diffuse emission from several carbon monoxide (CO) $J \rightarrow (J-1)$ rotational lines ($\nu_{\text{CO,}J} = 115 \cdot J$~GHz) for $J=4-7$ at redshifts $z<1$ and singly-ionized carbon ([CII]) in the $158.7$~\textmu m ($\nu_{\text{[CII]}}=1.889$~THz) line at redshifts $z=2.5-3.5$.\cite{Switzer2021_JATIS_EXCLAIM, Pullen2023_MNRAS_EXCLAIM} The EXCLAIM survey will consist of a 320~deg$^2$ extragalactic (EG) survey and several ${\sim}100$~deg$^2$ galactic plane (GP) survey regions. The GP survey will map CO(4-3) and neutral carbon ([CI]). The [CI] survey will assess the CO-H$_2$ ratio to inform the higher-redshift survey, as [CI] traces H$_2$ in regions where CO is photodissociated\cite{2015ApJ...811...13B, 2010ApJ...716.1191W}. The EG survey will be cross-correlated with galaxies and quasars in the Baryon Oscillation Spectroscopic Survey (BOSS)~\cite{2020arXiv200708991E} to map EXCLAIM observations in redshift space. 

EXCLAIM consists of a completely cryogenic telescope in an open liquid-helium bucket dewar with a cold inner volume approximately 1.5~m in diameter and 2.0~m deep. At target float altitudes above $27$~km, the ambient pressure of less than $10$~Torr pumps on the helium bath, lowering its temperature below its superfluid transition to approximately 1.7~K. Superfluid helium pumps demonstrated on the ARCADE2 and PIPER payloads efficiently cool the full optical chain to below 5~K.\cite{ARCADE2011_Singal, PIPER2016_Gandilo, Superfluid_Pumps_Kogut}

  \begin{figure*} [t]
  \begin{center}
  \begin{tabular}{c} 
  \includegraphics[clip=true, trim=0.5in 1.6in 0.6in 1in, width=1.0\textwidth]{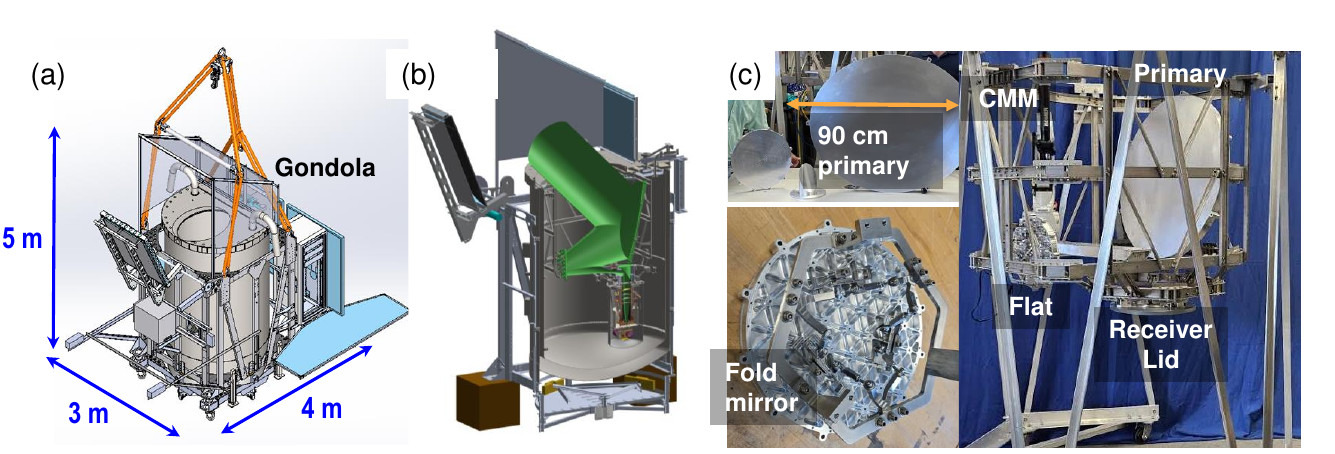}
  \end{tabular}
  \end{center}
  \caption[example] 
  { \label{fig:overview} 
Overview of the EXCLAIM optics within the EXCLAIM payload. Panel (a) shows the overall gondola with dimensions. The EXCLAIM telescope looks out through the aperture at the top of the cylindrical dewar. Panel (b) shows a cross-sectional view of the payload with the optical highlighted in green. Panel (c) shows individual photographs of the reflective optics (top left), the mounting structure of the folding flat mirror (bottom left) and the reflective optics mounted in the telescope frame along with the coordinate measuring machine (CMM) arm used to measure the relative positions of the optical components during alignment. }
  \end{figure*} 

As shown in Figure~\ref{fig:overview}, the EXCLAIM optics consist of a two-mirror, off-axis Gregorian telescope with a 90-cm parabolic primary mirror and a 10-cm parabolic secondary mirror. A folding flat between the primary and secondary allows the telescope to fit within the dewar volume. Baffling around the intermediate focus controls stray light and houses an aerogel scattering filter for rejection of IR radiation.~\cite{Aerogel_Filters_2020, Barlis2024_Aerogel_filters} The rays are collimated after the secondary mirror and pass through a meta-material anti-reflection (AR) coated silicon vacuum window into the receiver cryostat.\cite{EXCLAIM_Receiver_SPIE2024} The receiver houses additional stray-light baffling, aerogel scattering and band-defining metal-mesh filters,~\cite{Metal_Mesh_Review_2006} an AR-coated silicon lens,~\cite{Datta_Si_AR_2013} and six integrated silicon spectrometers (\textmu-Spec).~\cite{uSpec_Cataldo_2014, uSpec_Noroozian_2015, uSpec_Barrentine_2016, Cataldo_uSpec_2019}

The \textmu-Spec spectrometers combine all the elements of a traditional diffraction-grating spectrometer in a compact package using planar transmission lines on a silicon chip. The grating is replaced by a niobium microstrip delay line network that launches signals into a 2D parallel-plate waveguide region with emitting and receiving feeds arranged in a Rowland configuration.\cite{Rowland_1883} EXCLAIM will operate at the second grating order selected by an on-chip filter along with the free-space band-defining filters. A set of 355 kinetic inductance detectors (KIDs) made from 20-nm-thick aluminum (Al) half-wave resonators will detect the signal. A dipole slot antenna coupled to a hyper-hemispherical AR-coated silicon lenslet forms the spectrometer beam and couples the light to the on-chip circuitry. The combination of quasi-optical metal-mesh filters, on-chip filters, and spectrometer design defines the EXCLAIM observing band of $420$--$540$~GHz. The KID arrays for the six spectrometers use a microwave multiplexing readout scheme with heritage from The Next Generation Balloon-borne Large Aperture Submillimeter Telescope (BLAST-TNG)~\cite{BLAST_Lourie_2018} and the Far Infrared Observatory Mounted on a Pointed Balloon (OLIMPO).~\cite{OLIMPO_Presta_2020}

The EXCLAIM gondola design, attitude determination and control system, and flight electronics are based on those of PIPER with modest modifications and improvements. EXCLAIM will observe at constant elevation with scans in azimuth between $5^{\circ}$ and $10^{\circ}$ wide to map strips on the sky. A primary target is Sloan Stripe 82; however, the details of the scan strategy will depend upon the availability of sources for a given flight location and time of year. 


  \begin{figure*} [t]
  \begin{center}
  \begin{tabular}{c} 
  \includegraphics[clip=true, trim=0in 0in 0in 0in, width=0.9\textwidth]{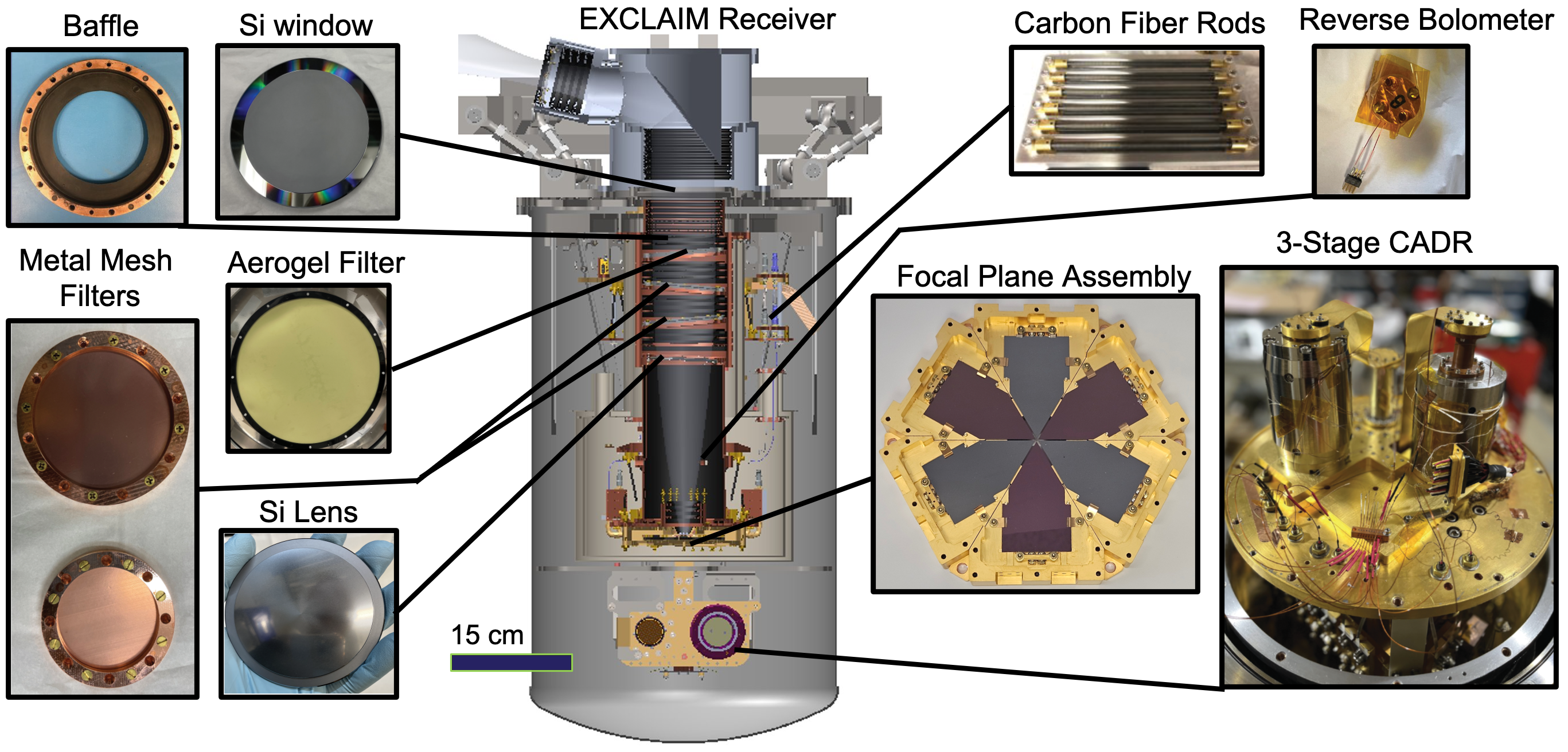} 
  \end{tabular}
  \end{center}
  \caption[example] 
  { \label{fig:render} 
Overview of the EXCLAIM receiver with key components indicated by photographs of fabricated elements. The secondary mirror sits at the top in a blackened enclosure with the intermediate focus at its left in this view surrounded by a set of baffles. The reverse bolometer is used for calibration of responsivity drifts during flight and is referred to elsewhere as the in-situ calibrator.}
  \end{figure*} 

This paper expands upon Ref.~\citenum{Exclaim_Optics_SPIE2022} and is organized as follows. Section~\ref{sec:optical_design} describes the optical design requirements, system layout, and key optical components. Section~\ref{sec:performance} describes the performance of the optical system as modeled with ray-tracing software and physical-optics analysis of diffraction in the system. Section~\ref{sec:optical_components} describes details of the fabrication of the reflective optics and other key components. We conclude in Section~\ref{sec:conclusion}.

\section{Optical Design}
\label{sec:optical_design}
\begin{figure} [t]
  \begin{center}
  \begin{tabular}{c} 
  \includegraphics[clip=true, trim=0in 0in 0in 0in, width=0.9\columnwidth]{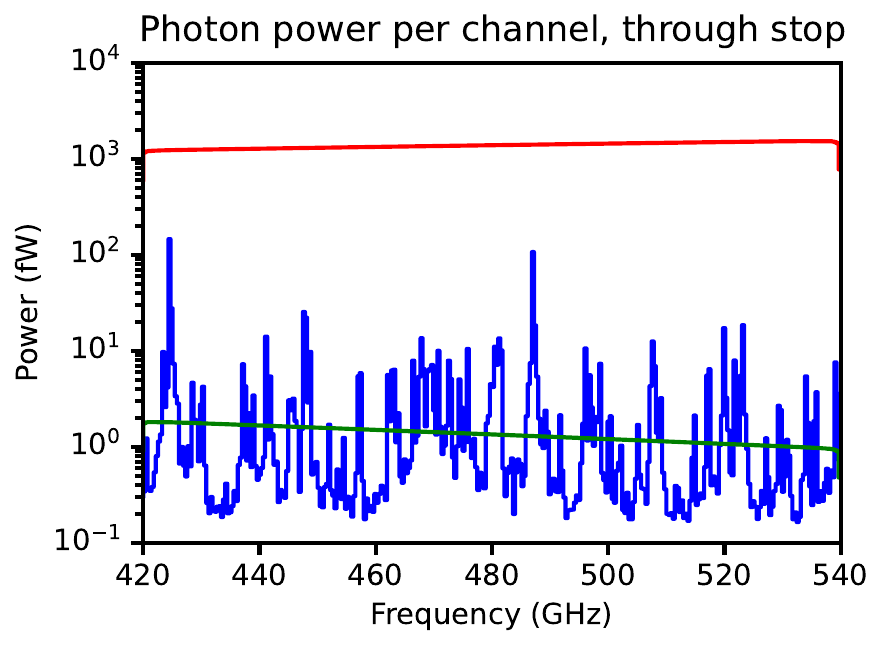}
  \end{tabular}
  \end{center}
  \caption[example] 
  { \label{fig:loading} 
Estimated optical power (defined as passing through the stop to the spectrometer) due to the atmosphere and a cold telescope (blue line) compared with blackbodies (100\% emissivity) at 5~K (green line) and 250~K (red line) for reference. The requirement on warm spill of $< - 40$~dB reduces the 250~K blackbody load below that of the estimated power from the atmosphere. The telescope temperature requirement of $< 5$~K ensures that the reflective optics with pessimistic emissivity of 10\% do not exceed the loading from the atmosphere in the darkest channels. }
  \end{figure}

The EXCLAIM optical design is unique in several ways. As EXCLAIM is not taking an instantaneous image and the map is filled in by multiple observations with different detectors in each map pixel, the optical design has relatively loose requirements on optical aberrations. There are, however, very stringent requirements on total excess loading on the detectors, since modest beam spill onto warm elements will significantly degrade performance in the darkest spectral channels. As shown in Figure~\ref{fig:loading}, even modest illumination at the $-40$~dB level of a 250~K surface is comparable to the power in the darkest channels. The EXCLAIM optical design aims to limit stray light through the overall system architecture, combined with baffling in key places. 

In this section, we describe the requirements on the optical system that flow down from EXCLAIM science and mission objectives (Section~\ref{sec:requirements_overview}); give an overview of the optical design that meets those requirements (Section~\ref{sec:optical_layout}); and describe the design of key optical components, including the silicon superfluid-tight receiver window, silicon lens, coupling to the spectrometer wafer via a dipole antenna and hyper-hemispherical silicon lenslet, IR-blocking aerogel scattering filters, metal-mesh band-defining filters, and in-flight calibration source (Section~\ref{sec:optical_components}).

\subsection{Requirements Overview}
\label{sec:requirements_overview}
The EXCLAIM optical design needs to satisfy a number of mechanical, optical, cryogenic, and integration and test (I\&T) requirements. Primary requirements on the optical design are:

\begin{enumerate}
    \item The telescope shall fit within a cylindrical volume approximately 1.2~m in diameter and 1.5~m deep. This corresponds to the usable volume within a bucket dewar sized such that the payload fits within mass limits set by balloon lift capabilities of $\sim 3400$~kg. The bucket dewar is the same size as has been demonstrated for the PIPER payload.\cite{PIPER2016_Gandilo} See Figure~\ref{fig:render} for more details on the payload mechanical design. 
    \item The angular resolution shall be $< 7^\prime$ to allow measurement of physical scales down to approximately 200~kpc (modes $k \gtrsim 5$~h/Mpc) for [CII] at $z \sim 3$.
    \item The design shall maintain total root-mean-squared (RMS) wave front error (WFE) below 0.075 waves at the center of the band, 480~GHz. This corresponds to a Strehl ratio greater than 0.80. 
    \item The instantaneous field of view (FOV) of the telescope shall be between $12.5^\prime$ and $25^\prime$. This allows efficient mapping of the target observation areas given the angular resolution.
    \item Excess loading from stray light on each KID shall be below the expected power in the darkest spectrometer channels, $\sim 0.1$~fW defined at the receiver cold stop, to ensure nearly background-limited performance across the band.
    \item The optical design shall provide for a collimated region within the receiver volume with a 3:1 aspect ratio for effective magnetic shielding, as the KIDs are susceptible to magnetic fields. 
    \item The telescope shall operate at $<5$~K to ensure that loading from the reflective optics and spill onto baffling, captured as a pessimistic 10\% total emissivity, does not exceed 0.1~fW, requiring that warm alignment translates to alignment upon cooling.
    \item The receiver shall be placed vertically under the primary mirror to allow the receiver to be tested in a liquid helium dewar prior to integration with the telescope and to allow the telescope to be aligned separately from and before integration with the receiver.  
    \item The design shall provide a means for rapid optical modulation in the event that the stability of the detector data is not sufficient. KIDs and the associated readout system can have excess low-frequency, or $1/f$, noise due to two-level systems that would lead to striping in the maps. The primary science signal band for EXCLAIM is approximately 5--25~Hz, based on the current scan strategy. This requirement ensures that rapid modulation could be implemented if required after laboratory testing or the engineering flight to mitigate excess $1/f$ noise.
\end{enumerate}

  \begin{figure*} [t]
  \begin{center}
  \begin{tabular}{cc} 
  \includegraphics[clip=true, trim=1.5in 0.05in 1.5in 2.8in, width=0.45\textwidth]{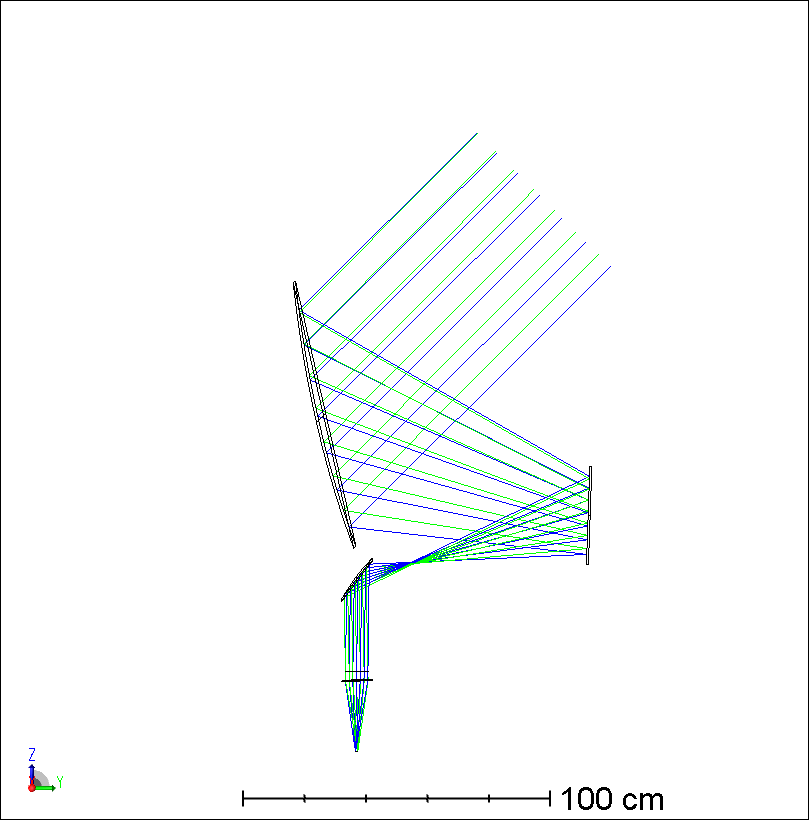} &
  \includegraphics[clip=true, trim=1.5in 2.8in 1.5in 0in, width=0.5 \textwidth]{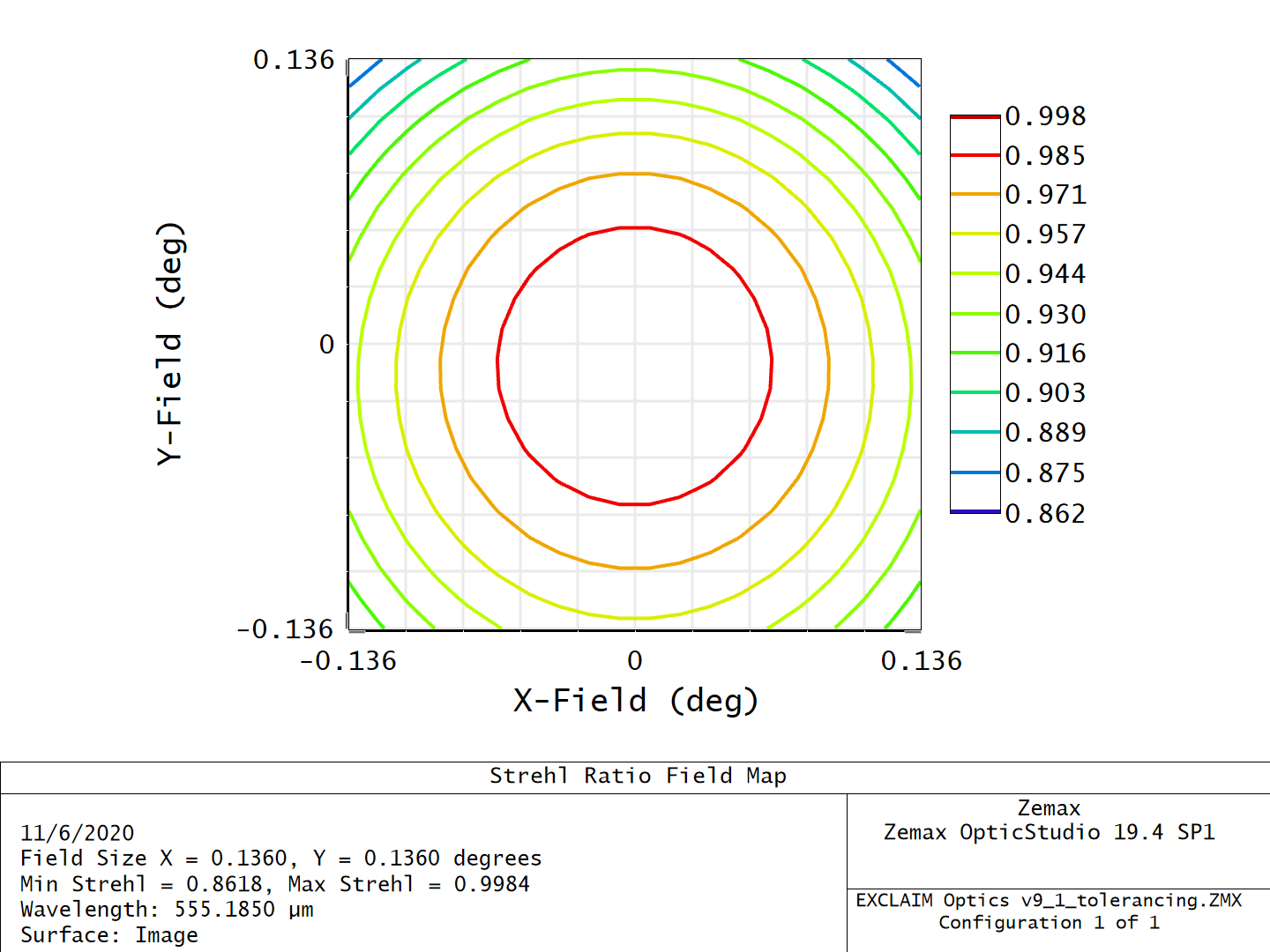}
  \end{tabular}
  \end{center}
  \caption[example] 
  { \label{fig:ray_trace} 
\textit{Left:} Ray trace of EXCLAIM optics using Zemax OpticStudio\footnote{Ansys Zemax OpticStudio, https://www.zemax.com/, Southpointe 2600 Ansys Drive
Canonsburg, PA 15317 USA} for two pixels at the top and bottom of the focal plane. \textit{Right:} Strehl ratio at the high end of the EXCLAIM band, 540~GHz, over the field of view, which for EXCLAIM is $16.1^\prime$, or $\pm 0.14^\circ$. The nominal design performance has Strehl ratio $>0.94$ across the field of view. Strehl ratio performance is better at lower frequencies within the band.}
  \end{figure*} 
  
These primary requirements lead to a series of derived requirements on the optical design. Key derived requirements are: 

\begin{enumerate}
  \setcounter{enumi}{0}
  \item The telescope shall have a physical aperture of 90~cm and projected aperture of 76~cm. This fulfills the angular resolution requirement while fitting within the allowable volume and allowing under-illumination of the primary mirror to reduce ambient-temperature spill. See Sec.~\ref{sec:diffraction} for more discussion of angular resolution given realistic Gaussian beam truncation. 
  \item The reflective telescope shall be $\sim f/2$ to stay within the allowable volume, while producing a collimated region inside the receiver of reasonable diameter, $<10$~cm, to provide the required 3:1 aspect ratio for magnetic shielding. 
  \item The receiver shall house a 1.8~K cold aperture stop that truncates the beams from the on-chip spectrometer lenslets at $<-15$~dB at all frequencies within the EXCLAIM band to maintain low edge illumination on the primary mirror for stray light reduction. Given the lenslet beam width (See Sec.~\ref{sec:optical_layout}), this requires that the system be $f/3.2$ after the silicon lens. 
  \item Stray light spill onto ambient temperature ($> 100$~K) surfaces, including those behind the primary mirror, around the dewar aperture, and the balloon, shall be less than $-40$~dB. This ensures that excess loading from warm spill stays below 0.1~fW.
\end{enumerate}

Though not a strict requirement, the design also attempted to reduce the potential for optical cavities to create fringing within the EXCLAIM band and multiple images of objects on the sky from ghosting. As a spectrometer, EXCLAIM is more prone to this than broadband imaging experiments. Given the design spectral resolving power, $R=512$, at the center of the EXCLAIM band, $\nu=470$~GHz, each KID channel is sensitive to a bandwidth of approximately $\Delta \nu=0.9$~GHz. This leads to a coherence length, $c/\Delta \nu$, of approximately $33$~cm, which is on the order of the size of the optics tube within the receiver. To mitigate possible cavities formed within the receiver optics tube, the silicon lens and filters were tilted with respect to the chief ray by $3^\circ$ and $2^\circ$, respectively. Different tilt directions were used for each optical element to further reduce cavity modes.

  \begin{figure*} [t]
  \begin{center}
  \begin{tabular}{cc} 
  \includegraphics[clip=true, trim=0in 0in 0in 0in, width=0.47\textwidth]{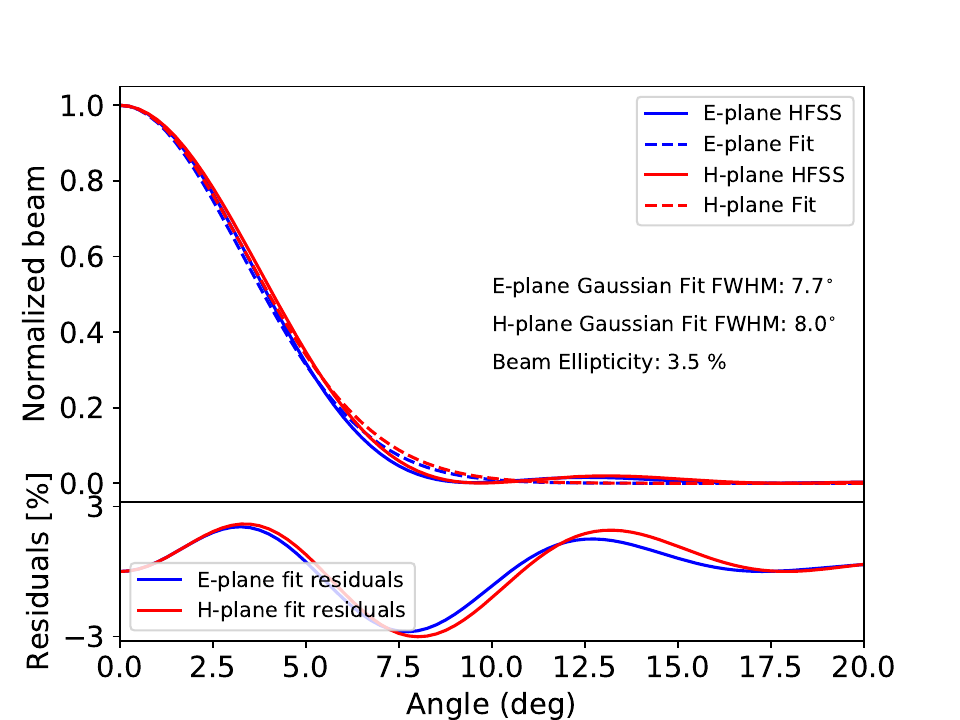} &
  \includegraphics[clip=true, trim=0in 0in 0in 0.3in, width=0.47 \textwidth]{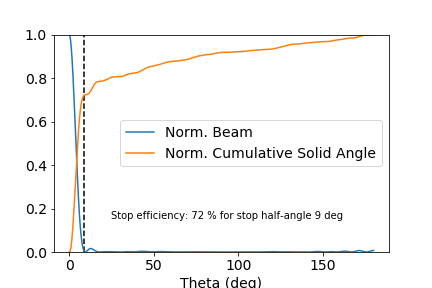}
  \end{tabular}
  \end{center}
  \caption[example] 
  { \label{fig:lenslet_beam} 
\textit{Left:} The beam simulated using Ansoft HFSS\footnote{Ansys HFSS, https://www.ansys.com, Southpointe 2600 Ansys Drive
Canonsburg, PA 15317 USA} for the EXCLAIM spectrometer on-chip silicon lenslet-coupled antenna at the center of the band (480~GHz). In the upper panel, the E-plane (H-plane) beam from HFSS is shown in solid blue (red) with a Gaussian fit shown in the blue (red) dashed lines. Residuals from the fit, shown in the lower panel, are below 3\% of the beam peak. \textit{Right:} Normalized beam in blue and cumulative solid angle in orange from HFSS simulations. The aperture efficiency for the EXCLAIM aperture stop half-angle of $9^\circ$ is 74\%.}
  \end{figure*}

\subsection{Optical Layout}
\label{sec:optical_layout}

To meet the requirements set out above, the EXCLAIM optical design shown in Figure~\ref{fig:ray_trace} uses a large, 90~cm, off-axis parabolic primary mirror that is 76~cm in projection. The primary mirror is highly under-illuminated to reduce beam spill around the edges of the primary, which could fall onto warm surfaces. The effective focal length of the primary mirror is set to 155~cm to allow the rays to come to an intermediate focus within the dewar volume on the way to the secondary mirror. The intermediate focus also provides a natural place for a chopper if rapid optical modulation is deemed necessary. A folding flat 30~cm in diameter redirects the rays so that the optical system can fit within the dewar and to allow the secondary mirror to be under the primary mirror, as required by the integration and test sequence. 

A more complicated design including a hyperbolic mirror in place of the folding flat in a Mizugutch-Dragone configuration~\cite{DragoneHogg1974, Mizugutch1976} was initially considered; however, the performance of the simpler design outlined here was found to be sufficient given the reasonably loose requirements on optical aberrations and cross polarization for EXCLAIM. Having a folding flat in place of a shaped mirror significantly reduces system complexity and mirror fabrication costs with modest loss of optical quality. 

  \begin{figure} [t]
  \begin{center}
  \includegraphics[clip=true, trim=3in 3in 3in 3in, width=0.4 \textwidth]{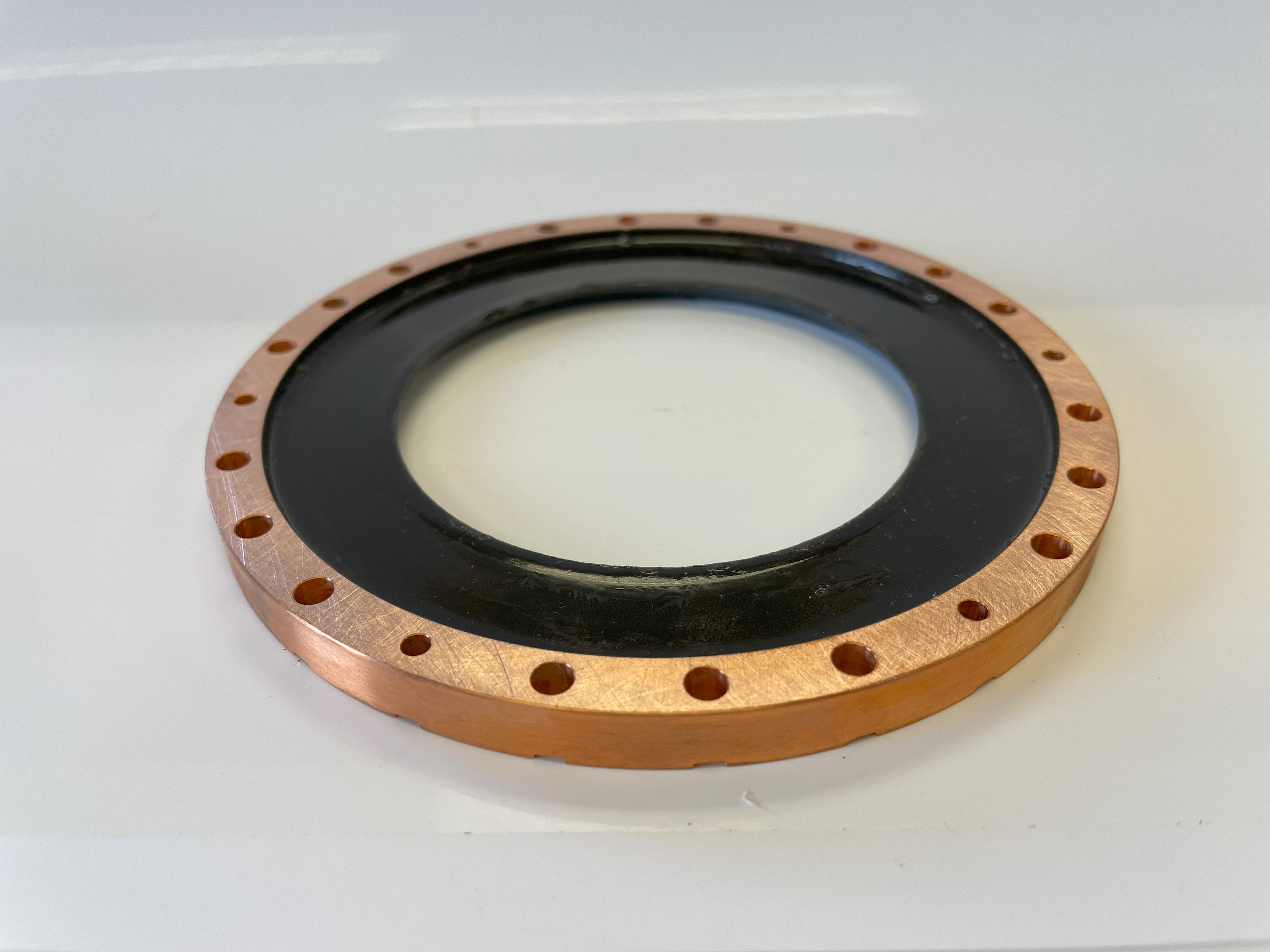}
  \end{center}
  \caption[example] 
  { \label{fig:baffle} 
Photograph of blackened EXCLAIM baffle for the optics tube.}
  \end{figure} 

The receiver houses a blackened 7.6~cm cold aperture stop and a plano-convex silicon lens with a focal length of 24~cm to couple light onto the hyper-hemispherical lenslets of the spectrometers. As noted above, the silicon lens is tilted by $3^\circ$ to reduce cavity modes and ghosting within the receiver. The planar surface of the tilted lens faces toward the focal plane. The 4-mm diameter lenslets provide an approximately Gaussian beam with a full-width at half maximum (FWHM) of $8.0^\circ$ (mean of E- and H-plane) at the center of the EXCLAIM band, 480~GHz. The corresponding $-15$~dB angle for the beams at that frequency is $15.8^\circ$ full angle. See Fig.~\ref{fig:lenslet_beam} for more details. At the low-frequency edge of the band at 420~GHz, the FWHM and $-15$~dB half angle are $9.0^\circ$ and $18.0^\circ$, respectively. As the beams are broadest for the lowest frequency within the band, requiring the cold stop to truncate at $<-15$~dB across the full band translates into a 7.6-cm cold stop diameter given the 24-cm lens focal length, leading to an optical speed of $f/3.2$ at the focal plane.

The six 4.0-mm diameter lenslets are equally spaced around a 9.0-mm circle (4.5~mm between neighboring lenslets) to give adequate space for the supporting silicon wafer and reasonable fabrication tolerances on the copper detector package. With the lenslet separation set, the field of view of the telescope is determined by the plate scale, which specifies the ratio between distance within the focal plane and angular separation on the sky. For this design, the plate scale is given by $F_s/(F_\ell F_p)$, where $F_p$, $F_s$, and $F_\ell$ are the focal lengths of the primary mirror, secondary mirror, and lens, respectively. Given the values above, the plate scale of the system is $1.8^\prime$/mm, corresponding to a field of view for the 9-mm focal plane of $16.1^\prime$, well within the range of Requirement~4. 


  \begin{figure*} [ht!]
  \begin{center}
  \begin{tabular}{cc} 
  \includegraphics[width=8.5cm]{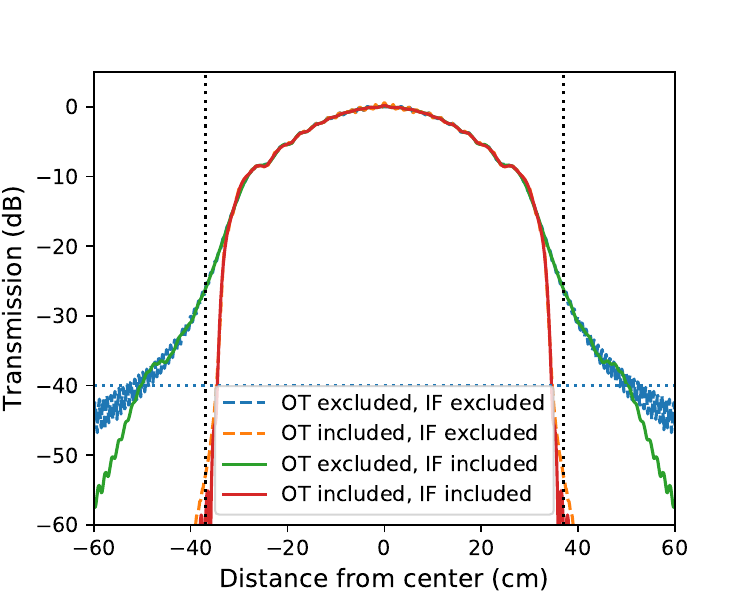} &
  \includegraphics[width=8.5cm]{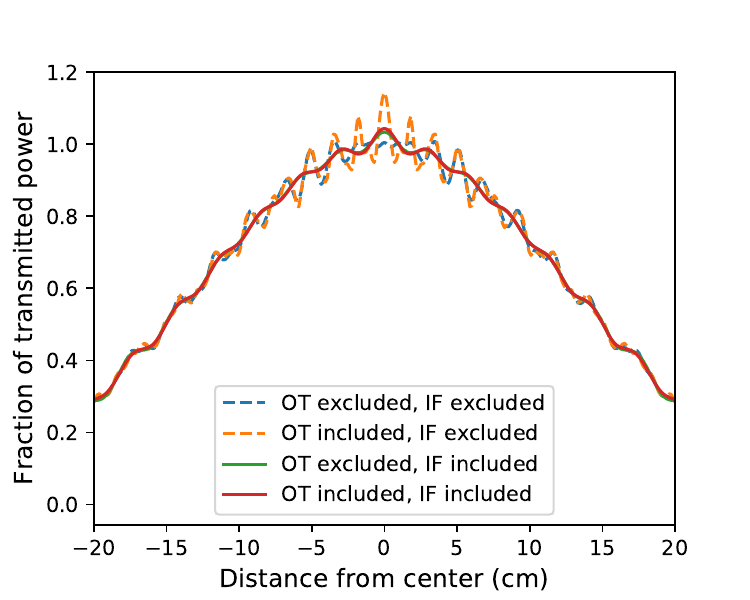}
  \end{tabular}
  \end{center}
  \caption{Illumination on the primary mirror.  We consider four different cases corresponding to the inclusion/exclusion of the optics tube (OT) baffles and the inclusion/exclusion of the intermediate focus (IF) baffle. \textit{Left}: Logarithmic scale with vertical dotted lines corresponding to the effective beam edges $38$ cm from the beam center, and the horizontal dotted line corresponding to the $-40$ dB level. Note that the two ``OT included'' curves are nearly indistinguishable on this plot. \textit{Right}: Linear scale highlighting diffraction effects on the main lobe. Here, the two `IF included' curves are nearly indistinguishable.}
  \label{fig:POPPY_primary_illum}
\end{figure*}

\subsection{Stray Light Control}
The intermediate focus is important for stray-light control, allowing the secondary mirror and receiver window to be placed entirely within a baffled enclosure at a temperature $\sim 1.7$~K, from which thermal radiation is negligible in the EXCLAIM band . In this enclosure and in the receiver optics tube, baffle rings are blackened with Epotek 377 epoxy loaded with silica and graphite powders.~\cite{Chuss_absorber} The thickness of this coating is chosen to be 2~mm to provide a balance between absorption at frequencies below the EXCLAIM band, mass, and heat capacity of the coating. The 2~mm coating is modeled using standard transfer matrix techniques to reflect no more than 20\% across the EXCLAIM band at typical angles of incidence $\sim 18^\circ$. The number of baffle blades and the separation between neighboring blades were chosen to provide a minimum of 9 reflections (attenuation of $\sim 0.2^9 = - 60$~dB) for a ray to escape in the specular reflectance limit. 

The dominant stray light path is likely to be scattering off baffle blade edges, including those of the cold stop. This stray-light path is strongly attenuated by having multiple baffle sections and enclosing the secondary mirror in a blackened cavity, which ensures that any scattered light from warm surfaces takes multiple bounces (minimum of 6 required for -40 dB attenuation) off blackened surfaces to reach a detector.

The thickness and shape of the coating is controlled either using molds following the procedures described in Ref.~\citenum{Blackening_Sharp_2012} or by filling pockets in the copper baffles, depending on the baffle. An example of a blackened baffle ring for the receiver optics tube is shown in Fig.~\ref{fig:baffle}.

After the intermediate focus, a parabolic secondary mirror with an effective focal length of 19.5~cm collimates the beam as it enters the receiver through the AR-coated silicon vacuum window. The collimated region is 23~cm long and serves multiple purposes. The initial driving requirement was to allow for magnetic shielding with a favorable 3:1 aspect ratio to be placed around the spectrometers, as the KIDs are sensitive to magnetic fields. Note that the magnetic shield diameter of 14~cm is necessarily larger than the diameter of the optical baffling to allow room for mechanical clearance between the magnetic shield and copper optics tube, as well as for mounting flanges for the lens, filters, and the optics tube as a whole. This means that the collimated region and focal length of the lens, at 24~cm, combine to provide an approximately 3:1 aspect ratio for the magnetic shielding. Optically, the collimated region affords an extended baffled section for stray light control and a natural location for free-space filters. 

\section{Modeled Performance and Tolerancing}
\label{sec:performance}

The EXCLAIM optical design was laid out and optimized using Zemax OpticStudio (Version 23.1.3) in sequential mode. Once the basic design described in Sec.~\ref{sec:optical_design} was determined, it was optimized to minimize wavefront error (WFE), allowing the lens radius and thickness to vary. Calculations of Strehl ratio, or equivalently WFE, across the field of view verify that the design is diffraction limited for all frequencies, as shown in Fig.~\ref{fig:ray_trace}. Below we describe possible sources of WFE and compile our best estimates of the contributions of each relative to the requirement in Table~\ref{tab:tolerance_stackup}.

After an optimized design was found, Zemax was used to place tolerances on important parameters in the system, including the lens radius and thickness, mirror tilts, and mirror displacements. As laid out in the requirements in Sec.~\ref{sec:requirements_overview}, the target WFE is 0.075 waves. The target mirror surface figures for the primary and secondary mirrors contribute 0.020 waves of WFE each, for a combined 0.028 waves when added in quadrature. This leaves 0.070 waves of WFE allowable for the misaligned system, where the nominal design has WFE of 0.040 waves. After an initial sensitivity analysis, in which individual parameters were varied alone to determine good ranges to explore for each, Monte Carlo (MC) simulations were performed using Zemax to determine tolerances for the system as a whole varying all parameters together. 

Tilts of the mirrors are the most tightly constrained, requiring that they be controlled to $\pm0.04^\circ$, $\pm0.1^\circ$, and $\pm0.4^\circ$ for the primary, folding flat, and secondary mirrors, respectively. Given the relative sizes of the mirrors, these tolerances require similar tolerances on the positions of the mount points on each mirror of approximately 0.5~mm each. Mirror decenter tolerances were found to be $\pm 1$~mm or greater, while tolerances on the relative distances of mirrors was found to be $\pm 3$~mm or greater. 

The telescope is held within the dewar using a stainless steel (SS) truss frame. The aluminum primary mirror and folding flat are attached to this frame via hexapod mounts with built-in flexures to allow for differential contraction upon cooling. Turnbuckles on each of the hexapod mounting arms provide precise control of the mirror positions. The aluminum secondary mirror is hard-mounted from the SS receiver lid with flexures to allow for differential contraction. A precision pinning arrangement will align the secondary mirror to the silicon lens and focal plane. Alignment points on the exterior of the secondary mirror allow the receiver and secondary mirror assembly to be aligned to the folding flat via a hexapod mount.

\subsection{Opto-Mechanical Design}
\label{sec:optomechanical}
A process has been developed to efficiently align the telescope, building upon PIPER heritage. A coordinate measuring machine (CMM)\footnote{Romer Absolute Arm, Model 7520, Hexagon Metrology, Oceanside, CA.} measures relative positions of the mirrors warm. These positions are compared with a model of the telescope when warm and hence slightly misaligned. Python code has been developed that provides adjustments to the hexapod mounting structure for the mirrors based upon CMM measurements using inverse kinematics. Initial tests of the alignment procedure have demonstrated the ability to align mirrors within $\pm 0.1$~mm in 6 iterations. This is well within the tolerance requirements that are typically $\sim \pm 0.5$~mm, as discussed in the previous section. Figure~\ref{fig:overview} shows the telescope being aligned in the flight frame.

The telescope is aligned warm with a calculated defocus to account for overall contraction of the SS telescope frame upon cooling to operating temperatures below 10~K. Values for the contraction of the SS frame and the Al mirrors are widely available in the literature (\textit{e.g.}, Ref.~\citenum{Pobell_Cryo_Book}) Fiducial values of fractional contraction ($\Delta l/l$) to low temperature of 0.30\% and 0.44\% are taken for SS and aluminum, respectively, giving differential contraction between the SS telescope frame and the Al mirrors of 0.14\%. Overall contraction of the telescope frame changes mirror center positions by $< 3$~mm, which can be modeled to well within the modest tolerances of $\pm 3$~mm on mirror-to-mirror distances discussed in the previous section. The flexure mounts holding the mirrors to the frame ensure that the positions of the mirror centers and their tilts remain nearly constant under differential contraction between the SS and Al. The primary mirror flexure mounts to a 40~cm diameter bolt circle that is offset perpendicular to the mirror surface by approximately 5~cm. Differential contraction of the frame and the mirror of 0.14\% amounts to $\sim 0.6$~mm that needs to be taken up by the flexure. Mis-estimation of differential contraction by as much as 10\% of the fiducial value (0.06~mm mis-alignment) would have negligible effect on the alignment of the telescope cold. The effects of thermal gradients are negligible, as all elements in the telescope cool well below a temperature of 5~K, based on PIPER experience,\cite{Superfluid_Pumps_Kogut}. At these low temperatures, residual thermal contraction is approximately two orders of magnitude below the contraction from room temperature to 5~K ($\Delta l/l < 3 \times 10^{-5}$), reducing misalignment and distortions in the mirror surfaces to negligible levels. In an absolute worst-case scenario, in which a gradient across the full 2~m span of the telescope causes a misalignment, this misalignment would be $< 6$~\textmu m.

Another possible source of wavefront error induced upon cooling is distortion of the mirror surface by the force of the flexure mounts for the mirrors as they deflect upon cooling. The stiffness of the flexure mounts was tuned to balance reduction of this distortion with the need to keep the overall structure stiff enough to properly constrain the mirrors. Finite-element analysis using Solidworks\footnote{Dassault Syst\`emes, https://www.solidworks.com, V\'elizy-Villacoublay, France} shows less than $\pm 5$~\textmu m distortion from this effect. Simulations were also carried out to evaluate the effect of gravitational distortion on the mirrors. This effect is most significant for the primary mirror, for which it is estimated to cause $\pm 15$~\textmu m. 

\begin{table*}[]
    \centering
    \begin{tabular}{c|c|c|c}
      \textbf{Element} & \textbf{WFE\footnote{Wavefront error} (waves)} & \textbf{\% of total\footnote{Contribution to variance when added in quadrature calculated as $100 \times (\sigma_j^2 / \sum_i{\sigma_i^2})$ for element $j$.}} & \textbf{Basis of Estimate} \\ \hline 
      Nominal Design                & 0.040 & 32\% & Zemax OpticStudio Ray Trace \\ \hline
      Alignment Tolerances (Worst-case $\pm 0.5$~mm)  & 0.053 & 57\%& Zemax OpticStudio Tolerancing Analysis \\ \hline
      Primary Mirror Surface Figure    & 0.010 & 2.0\% & Measured Actuals \\ \hline
      Folding Flat Surface Figure      & 0.003 & 0.2\% & Measured Actuals \\ \hline
      Secondary Mirror Surface Figure  & 0.010 & 2.0\% & Measured Actuals \\ \hline
      Misalignment Upon Cooling      & 0.005 & 0.5\% & 10\% error on cold position\\ \hline
      Mirror Gravitational Sag       & 0.012 & 2.9\% & Solidworks Simulation (Sec.~\ref{sec:optomechanical})\\ \hline
      Thermal Gradients              & 0.005 & 0.5\% & Worst-case 5~K gradient (Sec.~\ref{sec:optomechanical})\\ \hline
      Mechanical Stress from Flexure & 0.004 & 0.3\% & Solidworks Simulation (Sec.~\ref{sec:optomechanical}) \\ \hline
      Anisotropic Mirror Contraction & 0.008 & 1.3\% & Dilatometry (See Sec.~\ref{sec:mirror_manufacturing}) \\ \hline
      \textbf{Total} & \textbf{0.070} & \textbf{100\%}
    \end{tabular}
    \caption{Contibutions to total wavefront error (WFE) in wavelengths at the high end of the EXCLAIM band at 540~GHz. The most significant contributors are WFE of the nominal design and warm alignment tolerances. The requirement of 0.075 waves total WFE is still met with worst-case warm misalignment of up to $\pm 0.5$~mm. Achieved tolerances in test alignments suggest that alignment tolerances will be $\pm 0.2$~mm or less, providing significant margin of the design relative requirements.}
    \label{tab:tolerance_stackup}
\end{table*}



  \begin{figure*} [t]
  \begin{center}
  \begin{tabular}{cc} 
    \includegraphics[width=8.1cm]{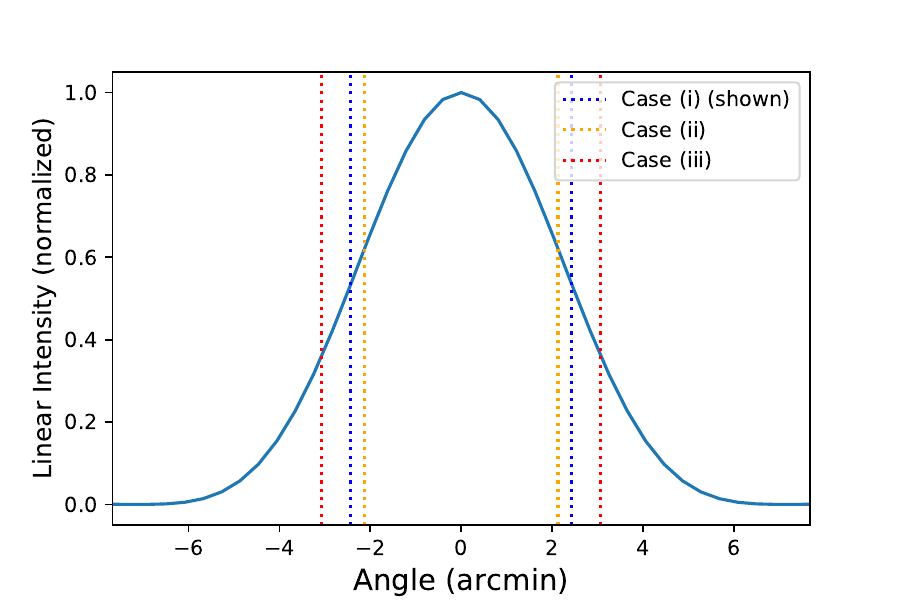} &  \includegraphics[width=8.1cm]{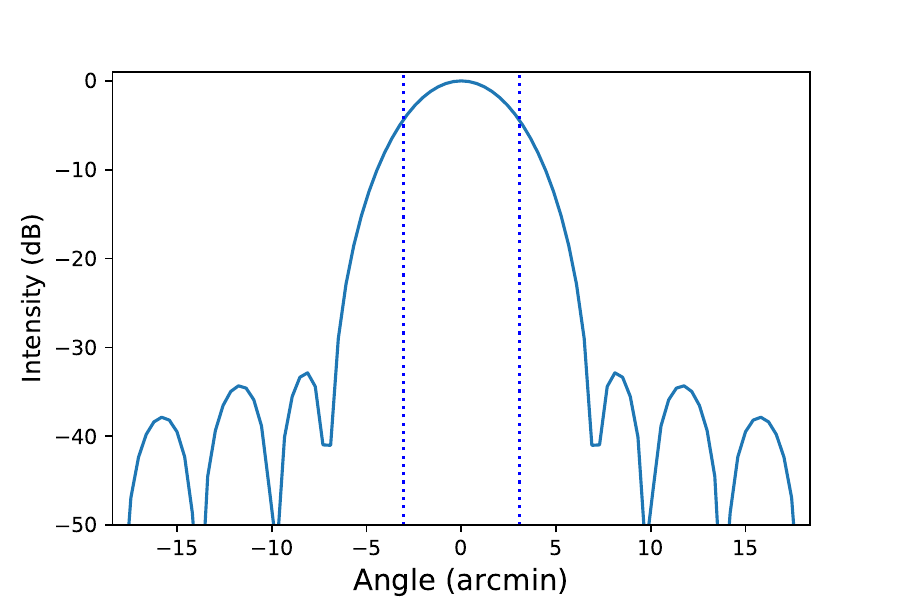} 
  \end{tabular}
  \end{center}
  \caption[example] 
  { \label{fig:POPPY_PSF} 
Far-field beam. \textit{Left}: Radial cut through the beam center with a linear scale. The three vertical lines correspond to the three different $\Theta_{\rm FWHM}^{\rm far-field}$ cases: (i) (blue) full POPPY simulation (shown); (ii) (orange) estimate based on the primary FWHM $\Theta_{\rm FWHM}^{\rm primary}$; (iii) (red) estimate based the primary edge taper $T_E (\rm dB)$ \textit{Right}:  Radial cut through the beam center with a log scale. The vertical line corresponds to Case (i).}
  \end{figure*}

\subsection{Diffraction Analysis}
\label{sec:diffraction}

Due to the stringent requirements on warm beam spill for EXCLAIM, a diffractive analysis of the optics was carried out to complement the ray-tracing results presented above. The primary goal of the diffraction analysis is to ensure that the spill on the primary is less than -40~dB, thereby minimizing loading from warm sources on the instrument.  The diffraction analysis is performed in two steps: (i) on-axis quasi-optics simulation via POPPY\footnote{This research made use of POPPY, an open-source optical propagation Python package originally developed for the James Webb Space Telescope project~\cite{perrin2012simulating}.}, and (ii) off-axis simulation via the Zemax physical optics module. The former is computationally inexpensive and allows us to simulate the full baffling assembly, though it does not include off-axis effects from the mirrors. The Zemax physical optics model offers a complementary approach, allowing accurate modeling of the effects of the off-axis mirror, but not detailed simulation of the effects of the baffles.

\begin{table*}[t]
    \centering
    \caption{Simulation parameters for POPPY simulation using the \texttt{Fresnel Optical System} setting.  The aperture at the lens is simulated as a Gaussian. The mirrors are simulated through POPPY as lenses with focal lengths denoted by $f$.  The baffles and the cold stop are simulated as pupils with radii given by $r$. Note that we consider 9 optics tube baffles, each with the same radius of truncation, and we do not consider the finite sizes of the lenses/mirrors because they contribute negligibly.}
    \begin{tabular}{c|c|c}
        \textbf{Element} & \textbf{Distance from cold stop} & \textbf{Parameters} \\
        \hline
        Lens aperture & 0 & FWHM $ = 2.302$~cm\\
        Cold stop & 0 & $r = 3.810$~cm \\
        Optics tube baffles & (0.89, 1.78, 2.67, 5.46, 6.35, 7.24, 10.03, 10.92, 11.81) cm & $r = 3.874$~cm \\
        Secondary mirror & 33.06 cm & $f = 19.50$~cm \\
        Intermediate focus baffle & 54.00 cm & $r = 4$~cm \\
        Primary mirror & 188.06 cm & $f = 155.0$~cm \\
    \end{tabular}
    \vspace{10pt}
    \label{tab:POPPY_params}
\end{table*}

The parameters for the POPPY simulation are shown in Table \ref{tab:POPPY_params}.  The simulation begins at the lens, at which the wavefront is well approximated as a Gaussian profile truncated by the cold stop. Figure~\ref{fig:lenslet_beam} shows this to be accurate to within $\pm 2$\%. The mirrors are simulated as lenses with corresponding focal lengths and the baffles are simulated as pupils.  We used the \texttt{Fresnel Optical System} setting to obtain the results shown in Figure \ref{fig:POPPY_primary_illum}. Four different cases are shown: (i) baffles at both the intermediate focus (IF) and optics tube (OT); (ii) IF baffle only; (iii) optics tube baffles only; (iv) no baffles. These simulations were performed for an optical frequency in the middle of the EXCLAIM band at $480$~GHz.

As shown in Figure \ref{fig:POPPY_primary_illum}, the optics tube baffling is necessary and sufficient to reduce spill at the edge of the primary to well below the -40\,dB target. In the case that OT and IF baffles are included, the main lobe is nearly Gaussian with a FWHM $\Theta_{\rm FWHM}^{\rm primary} =$ 22.4~cm and edge taper $T_E({\rm dB}) = 85$\,dB. The edge taper is defined as the absolute value of the primary illumination at the effective mirror radius of 38~cm, and it is heavily suppressed through the inclusion of OT baffles as shown in the left figure.  The baffle at the intermediate focus smooths out diffraction wiggles in the main lobe, as shown in the right figure, which we attribute to the truncation of the beam at the cold stop. In either case, the diffraction wiggles in the main lobe disappear when propagating to the far field.

The far-field beam for the fully-baffled POPPY simulation is shown in Figure \ref{fig:POPPY_PSF}.  The two plots show a slice through the beam center in a linear scale (left) and logarithmic scale (right). The far-field calculations were performed for three different cases for the illumination on the primary mirror: (i) full POPPY simulation; (ii) Gaussian with $\Theta_{\rm FWHM}^{\rm primary} =$ 22.4~cm, truncated at the primary effective radius; (iii) Gaussian with $T_E({\rm dB}) = 85$~dB.

In Case (i) we calculated the far-field pattern $\Theta_{\rm FWHM}^{\rm far-field}$ by Fourier transforming the primary illumination shown in Figure \ref{fig:POPPY_primary_illum}, resulting in $\Theta_{\rm FWHM}^{\rm far-field}$ = 4.33 arcmin. For the latter two cases we calculated  $\Theta_{\rm FWHM}^{\rm far-field}$ using the phenomenological relation\cite{goldsmith1987radiation}

\begin{equation}
    \Theta_{\rm FWHM}^{\rm far-field} = \left [1.02 + 0.0135 T_E({\rm dB})\right ] (\lambda/D).
    \label{eqn:FWHM}
\end{equation}

\noindent Equation \ref{eqn:FWHM} assumes a Gaussian primary mirror, truncated at an edge taper of $T_E({\rm dB})$, where $\lambda$ is the wavelength and $D$ is the effective diameter of the primary. For Cases (ii) and (iii) the truncated Gaussian estimates $\Theta_{\rm FWHM}^{\rm far-field}$ were 3.78 arcmin and 5.44 arcmin, respectively. $\Theta_{\rm FWHM}^{\rm far-field}$ was underestimated in Case (ii) by 0.55 arcmin because the mirror edges were over-illuminated, roughly corresponding to the exclusion of OT baffles.  $\Theta_{\rm FWHM}^{\rm far-field}$ was overestimated in Case (iii) because the bright main lobe was underestimated by only considering the heavily-suppressed edge illumination. While the analytical estimates were both incorrect, the fully-simulated beam did exhibit the expected linear scaling with wavelength to a precision of one part in $10^{7}$, demonstrating strong agreement with the scaling in Equation \ref{eqn:FWHM}. Comparing Cases (i) and (ii) we see only a modest increase in $\Theta_{\rm FWHM}^{\rm far-field}$, demonstrating that the inclusion of OT baffles accomplishes our goal of decreasing the spill on the primary, while not causing an unacceptable degradation in the far-field performance of the telescope.

\begin{figure*}
    \centering
    \begin{tabular}{cc}
     \includegraphics[width=0.4\linewidth]{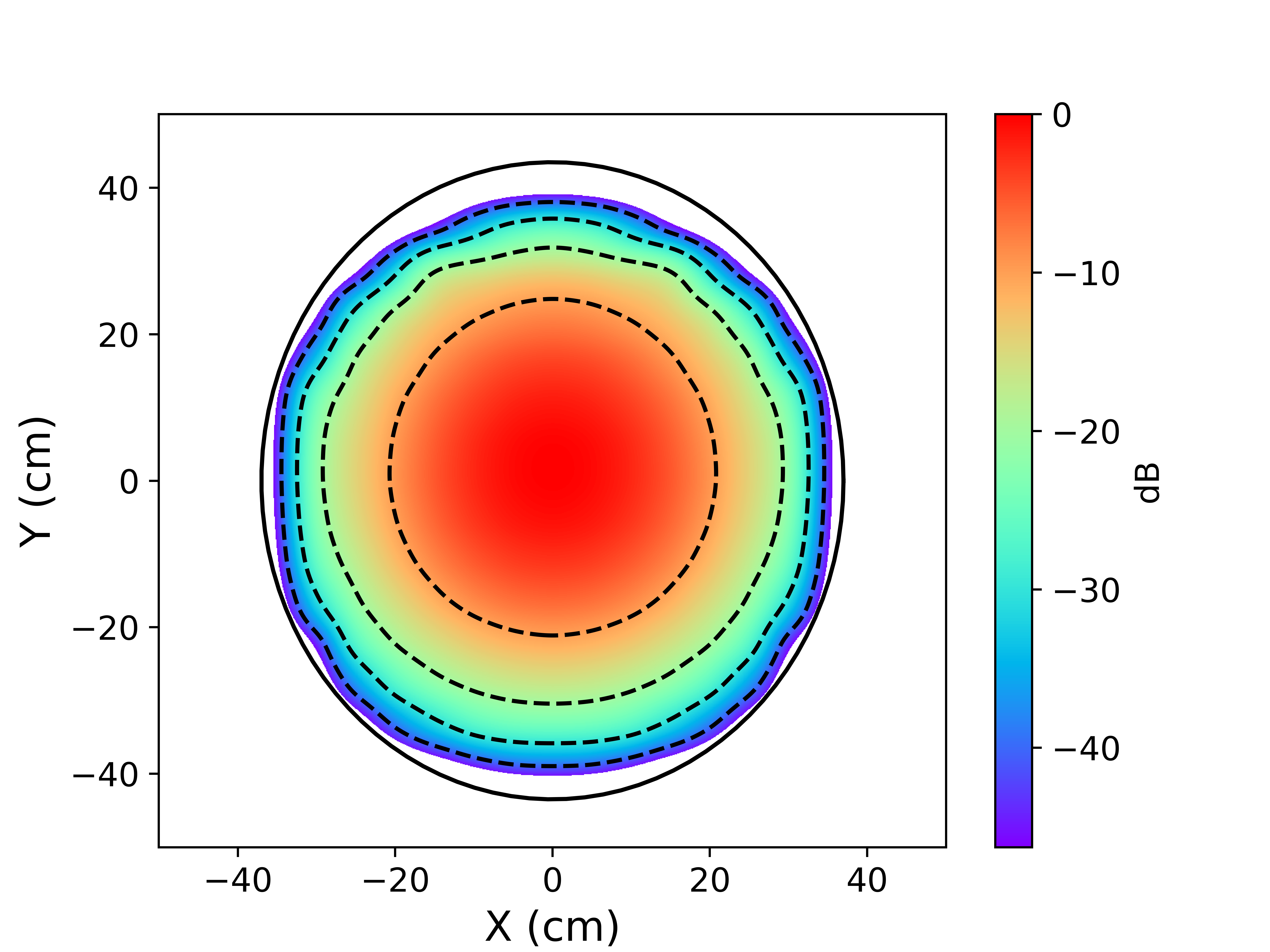} &
     \includegraphics[width=0.4\linewidth]{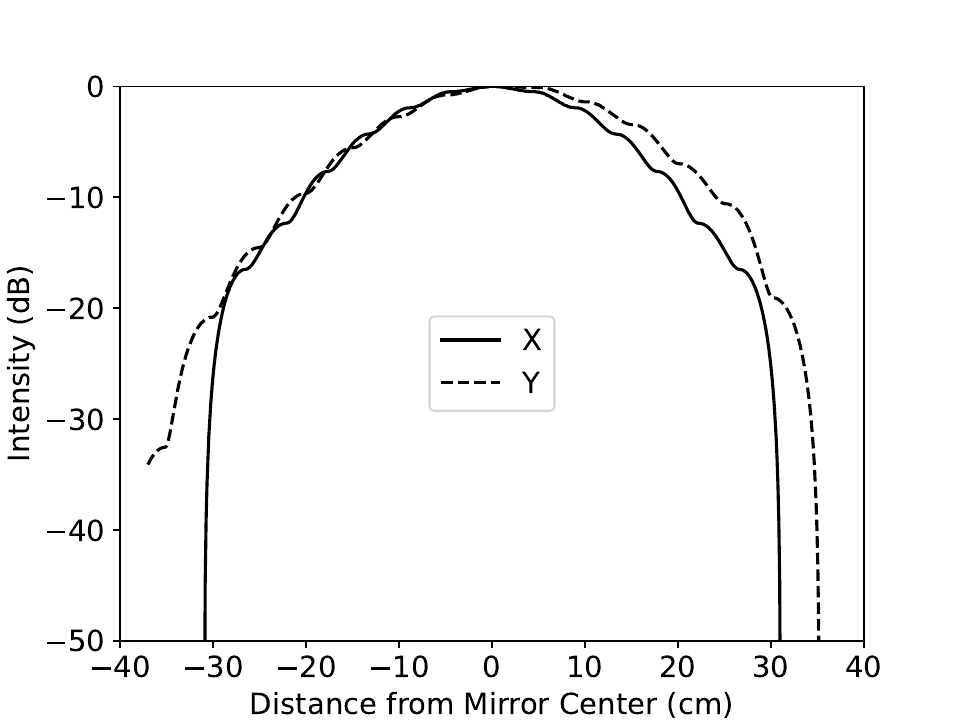} 
    \end{tabular}
    
    \caption{\textit{Top:} Zemax physical optics simulations of the illumination pattern on the EXCLAIM primary in dB relative to the peak amplitude. Dashed contour lines show -10, -20, -30, and -40 dB levels. The solid black ellipse shows the size of the main primary mirror surface. The Zemax simulation is truncated due to the edges of the secondary mirror and does not extend to the edges of the primary mirror, but reaches -45 dB, meeting instrument requirements of < -40~dB. \textit{Top:} X (horizontal) and Y (vertical) mirror illumination (in dB relative to the peak amplitude) pattern cuts. }
    \label{fig:zemax_physical}
\end{figure*}

Zemax physical optics (see Fig.~\ref{fig:zemax_physical} simulations of the off-axis system without accounting for the baffles show agreement with the POPPY simulations with edge taper on the primary mirror in excess of -50 dB. While the Zemax simulations do not include truncation due to the baffles, they by default truncate fields at the edges of the secondary mirror and folding flat. The combination of truncation at the cold stop and the secondary mirror edges effectively carry out much the same function as the baffles and lead to edge taper in excess of the required -40 dB. As the secondary mirror is enclosed in a $\sim 1.7$~K blackened cavity, beam spill around its edges will be terminated on cold surfaces and will not degrade sensitivity. 


\section{Optical Component Design and Manufacturing}
\label{sec:optical_components}

In this section we briefly summarize key design and manufacturing considerations for EXCLAIM optical components, including the aluminum primary, folding flat, and secondary mirrors; the silicon lens and receiver window, band-defining and IR blocking filters, and in-situ calibrator.

\begin{figure*}
\begin{center}
\includegraphics[clip=true, trim=0in 1.15in 0in 1in, width=0.95\textwidth]{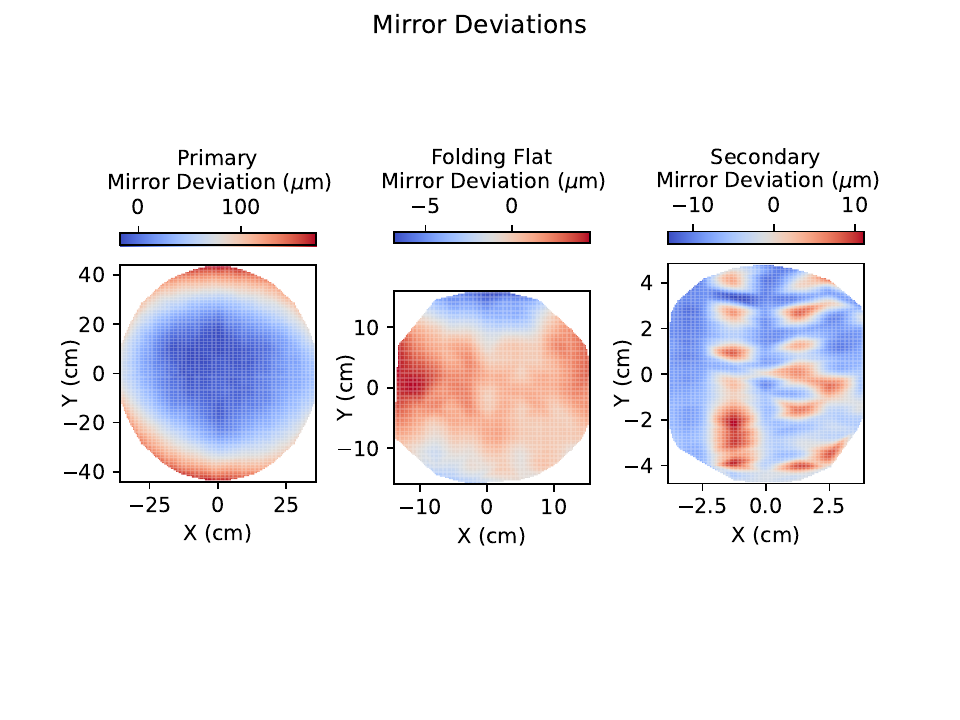}
\end{center}
\caption{\label{fig:mirror_deviations} 
Mirror deviations measured via coordinate measuring machine (CMM) for the primary mirror (\textit{left panel}), folding flat (\textit{center panel}), and secondary mirror (\textit{right panel}). The primary mirror RMS deviation within the inner 50~cm diameter where the majority of the EXCLAIM beam lies is 12~\textmu m. Deviations toward the circumference of the part, which functions to reduce warm spill in the sidelobes of the telescope optical response, approach 150~\textmu m, but will not significantly affect optical performance.}
\end{figure*}

\subsection{Mirror Manufacturing}
\label{sec:mirror_manufacturing}
The primary, folding flat, and secondary mirrors were computer numerical control (CNC) machined out of monolithic pieces of rolled aluminum 6061-T651\cite{Al_ASTM_B209} due to its combination of high strength and low stress. The large-scale surface figure of the mirrors is required to be within 25~\textmu m of nominal, corresponding to RMS phase error of 0.020 waves at 470~GHz. It was assumed in the design that the folding flat would be able to achieve considerably better surface figure, such that only contributions from the primary and secondary mirror surface figure errors, adding together in quadrature to give total mirror surface figure wavefront error less than 0.028 waves, were included in mirror requirements. As shown in Figure~\ref{fig:mirror_deviations}, achieved surface figures measured via CMM are well within this tolerance for the optically-important inner 50~cm diameter of the primary mirror (12~\textmu m RMS) and the entirety of the folding flat (6~\textmu m RMS) and secondary mirror (14~\textmu m RMS). Added in quadrature, these give a wavefront error of 0.016 waves at 470~GHz, achieving the 0.028 overall wavefront error allocated to the mirrors with 75\% margin.

The small-scale (sub-wavelength) RMS surface roughness of the mirrors is required to be below 0.4~\textmu m, so that scattering from the mirrors onto warm surfaces stays well below $-40$~dB. A surface roughness gives rise to a gain degradation factor of 

\begin{equation}
    \alpha = e^{-\left( 4 \pi \epsilon / \lambda \right)^2}
\end{equation}

\noindent where $\epsilon$ is the RMS surface roughness and $\lambda$ is the observing wavelength.\cite{Ruze_1966} Requiring this term to be below $- 42$~dB corresponds to an RMS surface finish of approximately 0.4~\textmu m. The delivered mirrors meet this requirement without the need for additional polishing. 

As the mirror surface figure is only measured at room temperature, while the telescope is at a temperature $< 5$~K during flight, there was concern about residual stresses in the material causing distortion of the mirror upon cooling. An uphill quench process was used to relieve stresses in the mirrors in between an initial rough cut of the mirror and the machining of the final mirror surface.\cite{Al_Quench_Simencio2011, Al_Quench_Mattos2017, Al_Quench_Robinson2022}

The thermal quench was performed by cycling the mirrors between a bath of liquid nitrogen (LN2) at approximately 77~K and a bath of boiling water at approximately 373~K. Each mirror underwent three full thermal cycles, which consisted of (1) lowering the mirrors into the LN2 bath and allowing them to equilibrate to $\sim77$~K, as indicated by the boiling of the LN2 bath becoming much less vigorous; (2) rapidly transferring the mirrors to the boiling water bath and waiting for the bath to return to a boil; and (3) rapidly transferring the mirrors to the LN2 bath. 

Rolled aluminum stock was chosen for the EXCLAIM mirrors over cast plates, as it was thought that rolled stock would provide better mechanical strength; however, there were concerns about anisotropy in the rolled material causing anisotropic contraction upon cooling that would distort the mirror surfaces. To address this concern, coupons were cut from each of the aluminum blocks used to make the primary and folding flat mirrors. Coupons were cut 6.3~mm in diameter by 25~mm long individually for the three orthogonal directions. Given the smaller size of the secondary mirror, distortion from anisotropic contraction was viewed as less important. Instead, two coupons were measured for each axis for the primary mirror to constrain the repeatability of the measurement and variations in anisotropy across the slab.

\begin{table*}
    \centering
    \caption{Measurements of the coefficient of thermal expansion (CTE) in parts per million (ppm) over the temperature range from room temperature to $- 110^\circ$C (163~K) using dilatometry for 6.3~mm diameter by 25~mm long coupons taken from the folding flat and primary mirrors. For each mirror, separate samples were taken from nearby areas for three orthogonal directions, labeled X, Y, and Z. For the primary mirror, two coupons were measured for each direction to constrain measurement errors and variations in CTE across the material. }
    \begin{tabular}{|c|c|c|c|c|c|c|c|c|c|} \hline
       &  \multicolumn{3}{c|}{Folding Flat} & \multicolumn{6}{c|}{Primary Mirror} \\ \hline
     Sample & X & Y & Z & X1 & X2 & Y1 & Y2 & Z1 & Z2 \\ \hline
     CTE (ppm) & 23.72 & 23.48 & 23.70 & 23.76 & 23.53 & 23.71 & 23.74 & 23.94 & 23.91 \\ \hline
    \end{tabular}
    \label{tab:dilatometry}
\end{table*}

As compiled in Table~\ref{tab:dilatometry}, the thermal contraction of each coupon was measured via dilatometry using a Netzsch 402 Expedis Supreme dilatometer\footnote{Netzsch, https://analyzing-testing.netzsch.com/en-US/products/dilatometry-dil} from room temperature to $-110^\circ$C (163~K), which was the limit of the machine's capability. Measurements showed linear contraction over this temperature range that was uniform to 0.2 parts per million (ppm), or 0.6\% of the average value of 23.7 ppm. If this level of variation were extrapolated to the $\sim 0.4$\% contraction ($\Delta l / l$) expected when cooling the material to 5~K,\cite{Pobell_Cryo_Book} it would contribute at most an additional 12~\textmu m deviation over the 50~cm diameter inner area the primary mirror, and far less for the much smaller secondary mirror. Added in quadrature with the surface deviations from measured at room temperature by CMM, this additional deviation from possible anisotropic contraction upon cooling would raise the RMS wavefront error from 0.016 waves to 0.018 waves. Even in a worst-case scenario in which the mirror anisotropic contraction errors added coherently with large-scale deviations in the room-temperature measurement to cause 24~\textmu m RMS deviations across the primary mirror upon cooling, this would only increase the combined wavefront errors from mirror deviations to 0.022 waves, which is still well within the 0.028 waves allocated to the mirrors.




\subsection{Silicon Lens and Receiver Window}
The receiver houses both a lens and a vacuum window, which are made of silicon. High-resistivity silicon exhibits low loss at sub-millimeter wavelengths.\cite{Lamb_1997, Datta_Si_AR_2013}

Due to differential thermal contraction between silicon and common laminate AR coating materials, the lens and window will feature a metamaterial AR coating. As described in Ref.~\citenum{EXCLAIM_Receiver_SPIE2024}, a two-layer coating consisting of nested sub-wavelength cuts are diced into the optic surface using a custom dicing saw.\cite{Datta_Si_AR_2013} This AR coating reduces reflections to the sub-percent level across the EXCLAIM bandpass. Silicon lenses with diced metamaterial AR coatings have been demonstrated and deployed on numerous millimeter-wave instruments such as the Atacama Cosmology Telescope and Simons Observatory.\cite{Thornton_ACTpol_2016, Golec_AR_Coating_2016} The AR coating for EXCLAIM will push the high-frequency limit for the metamaterial technology as it will be the widest bandwidth metamaterial AR coating to be deployed at near-THz frequencies.

EXCLAIM will demonstrate a superfluid-tight silicon vacuum window for the first time. The window must maintain high in-band transmission, while providing a superfluid-tight vacuum seal. PIPER has previously demonstrated a quartz window with a superfluid-tight indium seal and a polytetrafluoroethylene (PTFE) AR coating;\cite{PIPER_Window_Datta_2020} however, the measured quartz loss at EXCLAIM frequencies is significant, leading to the choice of silicon as an alternative. We have tested and confirmed that a silicon window with indium seal provides a superfluid-tight vacuum seal with flanges similar to the ones on the EXCLAIM receiver window.\cite{EXCLAIM_Receiver_SPIE2024} We plan to repeat that test with an AR coated window to confirm that the fabrication of the metamaterial coating does not compromise the superfluid-tight vacuum seal.


  \begin{figure*} [t]
  \begin{center}
  \begin{tabular}{cc} 
  \includegraphics[clip=true, trim=0in 0in 0in 0in, width=0.5\textwidth]{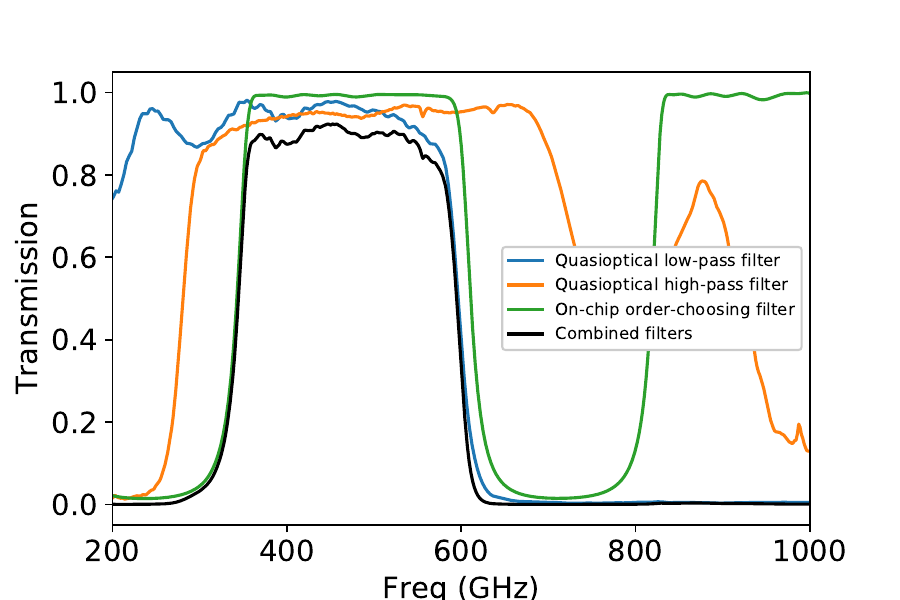} &
  \includegraphics[clip=true, trim=0in 0in 0in 0in, width=0.45 \textwidth]{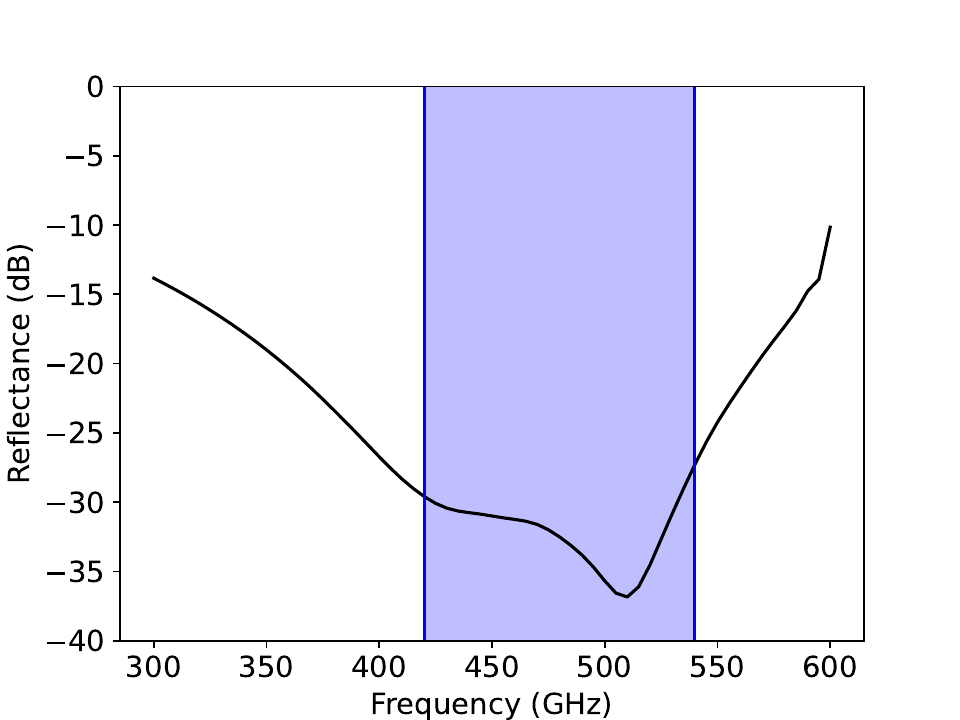}
  \end{tabular}
  \end{center}
  \caption[example] 
  { \label{fig:filters} 
 \textit{Left:} Compilation of quasioptical, metal-mesh low-pass and high-pass filters (housed in the receiver optics tube) and on-chip order-choosing filter planned for EXCLAIM, along with the estimated total transmission spectrum through the set of filters. \textit{Right:} Modeled performance of the baseline two-layer diced AR coating for the EXCLAIM silicon lens and window. The filled blue area indicates the EXCLAIM band (420-540 GHz).}
  \end{figure*} 



\subsection{Filters}
EXCLAIM requires free-space filters for IR rejection and band definition. Two IR-blocking filters will be used in the EXCLAIM system, one near the intermediate focus and the other as the first element after the vacuum window within the receiver volume. These IR blocking filters will be composed of diamond scattering particles embedded in a polyimide aerogel substrate with low index of refraction and low loss.~\cite{Aerogel_Filters_2020, Barlis2024_Aerogel_filters} Particle size and density is tuned to produce a lowpass filter with a cutoff frequency $\sim 1$~THz. 

Two free-space filters for band definition will be placed between the IR blocking filter and the aperture stop within the receiver optics tube and will be composed of a heat-pressed stack of metal-mesh filters on polypropylene film with dielectric spacers.~\cite{Metal_Mesh_Review_2006} A combination of a highpass and lowpass filter, used in conjunction with an order-choosing filter on the spectrometer wafer, will reject radiation immediately above and below the EXCLAIM band. Fig.~\ref{fig:filters} shows the modeled transmission spectra of the band-defining filters along with the resultant composite spectrum. 

\subsection{In-Situ Calibrator}
The EXCLAIM receiver will incorporate an in-situ calibration source for in-flight characterization of spectrometer response, uniformity, and time-varying responsivity. The source is a reverse bolometer of similar design to that used by the Herschel SPIRE instrument.~\cite{Calibrator_Pisano_2005} It consists of a sapphire square with an integrated heater thermally isolated from the bath stage using nylon threads. This reverse bolometer is mounted in a copper ring and placed within a bead-blasted integrating cavity. A small aperture in the cavity couples power out to the spectrometers. The calibrator is placed just below and outside the lens to couple into the sidelobes of the on-chip spectrometer lenslet beams.

\section{Conclusion}
\label{sec:conclusion}

The optical design for the EXCLAIM instrument has been described. This design meets all requirements placed on it by the science goals and overall mission architecture. The catadioptric system employs a 90~cm parabolic primary mirror, 30~cm folding flat, and 10~cm parabolic secondary mirror, along with a 10~cm silicon lens inside the receiver optics tube, to provide $4.2^\prime$ FWHM resolution in the center of the EXCLAIM band at 480~GHz over a $22.5^\prime$ field of view. 

The EXCLAIM mission design requires tight control over spill onto warm surfaces at the top of the dewar where light enters the cryogenic telescope. Particular care has been taken in the design to ensure that spill onto warm elements remains below $-40$~dB. This is achieved by under-illuminating the primary mirror at the expense of achievable angular resolution and through the use of absorptive baffling in the receiver optics tube and around the intermediate focus between the primary and secondary mirrors. Both analytical POPPY simulations in an on-axis analog system including apodization from baffling and physical optics simulations of the full off-axis optical system in Zemax OpticStudio without baffle apodization suggest that the current system meets this requirement with significant margin. 



  

\nocite{*}
\bibliography{exclaim-optics}

\end{document}